\documentclass[twocolumn,showpacs,preprintnumbers,amsmath,amssymb,prb]{revtex4}

\usepackage{graphicx}
\usepackage{bm}

\usepackage{amsfonts}

\newcommand{\Div}{\mathop{\rm div}\nolimits}
\newcommand{\curl}{\mathop{\rm curl}\nolimits}

\begin{document}
\title{Two-dimensional ring-like vortex and
multisoliton nonlinear structures at the upper-hybrid resonance}
\author{V. M. Lashkin}
\email{vlashkin@kinr.kiev.ua}
 \affiliation{Institute for Nuclear
Research, Pr. Nauki 47, Kiev 03680, Ukraine}

\date{\today}

\begin{abstract}
Two-dimensional (2D) equations describing the nonlinear
interaction between upper-hybrid and dispersive magnetosonic waves
are presented. Nonlocal nonlinearity in the equations results in
the possibility of existence of stable 2D nonlinear structures. A
rigorous proof of the absence of collapse in the model is given.
We have found numerically different types of nonlinear localized
structures such as fundamental solitons, radially symmetric
vortices, nonrotating multisolitons (two-hump solitons, dipoles
and quadrupoles), and rotating multisolitons (azimuthons). By
direct numerical simulations we show that 2D fundamental solitons
with negative hamiltonian are stable.
\end{abstract}

\maketitle

\section{Introduction}

Upper-hybrid (UH) waves are frequently observed in space and
laboratory plasmas. The UH waves can be excited by beam
instabilities, mode conversion of extraordinary electromagnetic
waves at the upper-hybrid resonance layer, etc. \cite{Shukla0}
One-dimensional (1D) theory of the nonlinear UH waves interacting
with the low-frequency motions of magnetosonic type was developed
in Refs. $2$--$7$. In particular, for low frequency
magnetohydrodynamic perturbations with frozen-in field lines and
for the negative group dispersive UH waves, Kaufman and Stenflo
\cite{Stenflo} showed the existence of UH solitons with
compressional density (magnetic field) perturbations in the
super-magnetosonic regime. Dispersive magnetosonic/lower-hybrid
waves interacting with high-frequency one-dimensional UH waves
were first considered in Refs. $4$ and $5$ and then were studied
in more detail, including intensive numerical modelling
\cite{Shukla3}.

The aim of this paper is to present two-dimensional (2D) theory of
nonlinearly coupled dispersive magnetosonic and high-frequency UH
waves. We have derived a set of corresponding nonlinear equations
and found numerically different types of 2D nonlinear localized
structures such as fundamental solitons, radially symmetric
vortices, nonrotating multisolitons (dipoles and quadrupoles), and
rotating multisolitons (azimuthons).

Dispersion of the magnetosonic wave effectively introduces a
nonlocal nonlinear interaction, that is the nonlinear response
depends on the wave packet intensity at some extensive spatial
domain. Nonlocal nonlinearity naturally arises in many areas of
nonlinear physics and plays a crucial role in the dynamics of
nonlinear coherent structures. While collapse is a usual
phenomenon in the multidimensional Zakharov-like models with local
low-frequency responce, nonlocal nonlinearity can arrest collapse
and lead to stable multidimensional localized structures
\cite{Yakimenko, Briedis, We1, Kartashov, Lopez, Skupin}. Turitsyn
proved \cite{Tur} the absence of collapse for three particular
shapes of the nonlocal nonlinear response in the multidimensional
generalized nonlinear Schr\"{o}dinger equation (GNSE). Later, a
rigorous proof of absence of collapse in arbitrary spatial
dimensions during the wave-packet propagation described by the
nonlocal GNSE with sufficiently general symmetric response kernel
was presented in Ref. $15$. In the present paper we give a proof
of absence of collapse for the 2D model describing the nonlinear
interaction between upper-hybrid and dispersive magnetosonic
waves.

The paper is organized as follows. In Sec.~\ref{sec2}, we present
the generalized Zakharov-type system of 2D equations, describing
the interaction between high-frequency upper-hybrid waves and
low-frequency dispersive magnetosonic waves. A linear stability
analysis is performed in Sec.~\ref{sec3}. Section~\ref{sec4}
contains the rigorous proof of absence of collapse in the model.
Localized nonlinear solutions, including vortex ring-like,
nonrotating (a monopole, a dipole, and a quadrupole) and rotating
(azimuthons) multisoliton solutions are presented in
Sec.~\ref{sec5}. The conclusion is made in Sec.~\ref{sec6}.

\section{Derivation of equations}
\label{sec2}

We consider a homogeneous electron–-ion plasma in a uniform
external magnetic field $\mathbf{B}_{0}=B_{0}\hat{\mathbf{z}}$,
where $\hat{\mathbf{z}}$ is the unit vector along the z direction.
In the linear approximation, the UH waves are characterized by the
dispersion relation
\begin{equation}
\omega=\omega_{UH}\left(1+\frac{1}{2}k_{\bot}^{2}R^{2}\alpha\right),
\end{equation}
where $\omega_{UH}=(\omega_{pe}^{2}+\omega_{ce}^{2})^{1/2}$ is the
UH resonance frequency, $\omega_{pe}$ ($\omega_{ce}$) is the
electron plasma (gyro) frequency, $v_{te}=(T_{e}/m)^{1/2}$ is the
electron thermal speed,
$\alpha=\omega_{pe}^{2}/(\omega_{pe}^{2}-3\omega_{ce}^{2})$, and
$R^{2}=3v_{te}^{2}/\omega_{UH}^{2}$. Note that the dispersion of
the UH waves is negative for $\omega_{pe}^{2}<3\omega_{ce}^{2}$.

Equation for the slow varying complex amplitude $\varphi$ of the
potential of the high-frequency electrostatic electric field
\begin{equation}
\mathbf{E}^{H}=-\frac{1}{2}[\nabla\varphi\exp(-i\omega_{UH}t)+\mathrm{c.c.}]
\end{equation}
of the upper hybrid wave can be obtained from the equation
\begin{equation}
\nabla\cdot(\hat{\varepsilon}\nabla\varphi)=0.
\end{equation}
with
\begin{equation}
\hat{\varepsilon}=\begin{pmatrix} \varepsilon_{\bot} & ig & 0 \\
-ig & \varepsilon_{\bot} & 0 \\
0 & 0 & \varepsilon_{||}
\end{pmatrix},
\end{equation}
where the dielectric tensor $\hat{\varepsilon}$ is considered as a
differential operator with $\omega\rightarrow
\omega_{UH}+i\partial/\partial t$ (assuming
$\omega_{UH}\gg\partial/\partial t$) and $\mathbf{k}\rightarrow
-i\nabla$. Under this, the nonlinear perturbations of the plasma
density $\delta n$ and magnetic field $\delta B$ are taken into
account in $\varepsilon_{\bot}$, $\varepsilon_{||}$ and $g$, so
that substitutions $n_{0}\rightarrow n_{0}+\delta n$ and
$B_{0}\rightarrow B_{0}+\delta B$ are made in the resulting
equation
\begin{equation}
\label{eps}
\nabla_{\bot}\cdot(\varepsilon_{\bot}\nabla_{\bot}\varphi)+\frac{\partial}{\partial
z }\left(\varepsilon_{||}\frac{\partial\varphi}{\partial
z}\right)+i\hat{\mathbf{z}}\times\nabla g\cdot\nabla\varphi=0.
\end{equation}
 As a
result, we have
\begin{gather}
\label{eq1}
\Delta\left(2i\omega_{UH}\frac{\partial\varphi}{\partial t
}+3v_{te}^{2}\alpha\Delta\varphi\right)+
\frac{\omega_{pe}^{2}\omega_{ce}^{2}}{\omega^{2}_{UH}}
\frac{\partial^{2}\varphi}{\partial z^{2}}\\ \nonumber
 =\nabla\cdot\left\{\left(\omega_{pe}^{2}\frac{\delta n}{n_{0}}+
 2\omega_{ce}^{2}\frac{\delta B}{B_{0}}\right)\nabla\varphi \right.
 \\ \nonumber
\left.
-i\frac{\omega_{ce}}{\omega_{UH}}\left[\omega_{pe}^{2}\frac{\delta
n}{n_{0}}+(\omega_{pe}^{2}+2\omega_{ce}^{2})\frac{\delta
B}{B_{0}}\right] \nabla\varphi\times\hat{\mathbf{z}}\right\}.
\end{gather}
The second term in the $\{\dots\}$ bracket in Eq. (\ref{eq1})
comes from the last term in Eq. (\ref{eps}) and corresponds to the
so-called vector nonlinearity. This term is identically zero for
1D case and for the fields with axial symmetry. This term can also
be neglected if $\omega_{ce}/\omega_{pe}\ll 1$.

The upper-hybrid  waves have wave numbers almost normal to the
external magnetic field ($k_{z}\ll k_{\bot}$) and, in the
following, we will consider two-dimensional (2D) case with
$k_{z}=0$ so that $\Delta=\partial^{2}/\partial
x^{2}+\partial^{2}/\partial y^{2}$ and $\nabla=(\partial/\partial
x,\partial/\partial y)$.

 The low-frequency motion of the plasma is governed by the
continuity and momentum equations for ions and electrons. We
assume quasineutrality condition, so that $\delta n_{i}=\delta
n_{e}\equiv \delta n$. Thus, we have
\begin{equation}
\label{ni} \frac{\partial \delta n}{\partial
t}+n_{0}\nabla\cdot\mathbf{v}_{i}=0,
\end{equation}
\begin{equation}
\label{vi} \frac{\partial \mathbf{v}_{i}}{\partial
t}=\frac{e}{M}\mathbf{E}-\frac{\gamma_{i}T_{i}}{n_{0}M}\nabla\delta
n+\omega_{ci}[\mathbf{v}_{i}\times\hat{\mathbf{z}}],
\end{equation}
\begin{equation}
\label{ne} \frac{\partial \delta n}{\partial
t}+n_{0}\nabla\cdot\mathbf{v}_{e}+\nabla\cdot \mathbf{F}_{2}=0,
\end{equation}
\begin{equation}
\label{ve}
 \frac{\partial \mathbf{v}_{e}}{\partial
t}+\mathbf{F}_{1}=-\frac{e}{m}\mathbf{E}-\frac{\gamma_{e}T_{e}}{n_{0}m}\nabla\delta
n-\omega_{ce}[\mathbf{v}_{e}\times\hat{\mathbf{z}}],
\end{equation}
where $\gamma_{i}$ ($\gamma_{e}$) is the ion (electron) ratio of
specific heats, and
\begin{equation}
\mathbf{F}_{1}=\langle
(\mathbf{v}^{H}\cdot\nabla)\mathbf{v}^{H}\rangle
+\left\langle\frac{e}{mc}[\mathbf{v}^{H}\times\mathbf{B}^{H}]\right\rangle,
\quad \mathbf{F}_{2}=\langle n^{H}\mathbf{v}^{H}\rangle
\end{equation}
are nonlinear terms in electron equations, the angular brackets
denote averaging over the high-frequency oscillations, and the
superscripts $H$ denote corresponding quantities for the
high-frequency fields. Multiplying Eq. (\ref{ve}) by $m/M$ and
adding with Eq. (\ref{vi}), we have
\begin{equation}
\label{ie}
 \frac{\partial}{\partial
t}\left(\mathbf{v}_{i}+\frac{m}{M}\mathbf{v}_{e}\right)=-\frac{m}{M}\mathbf{F}_{1}
-\frac{v_{s}^{2}}{n_{0}}\nabla\delta n-
\omega_{ci}[(\mathbf{v}_{e}-\mathbf{v}_{i})\times
\hat{\mathbf{z}}],
\end{equation}
where we have introduced the effective sound speed
$v_{s}=\sqrt{(\gamma_{i}T_{i}+\gamma_{e}T_{e})/M}$. Taking the
$\Div$ from Eq. (\ref{ie}) we get
\begin{gather}
\label{ie1}
 \frac{\partial}{\partial
t}\left(\nabla\cdot\mathbf{v}_{i}+\frac{m}{M}\nabla\cdot\mathbf{v}_{e}\right)
=-\frac{m}{M}\nabla\cdot\mathbf{F}_{1}\nonumber \\
-\frac{v_{s}^{2}}{n_{0}}\Delta\delta n-
\omega_{ci}\hat{\mathbf{z}}\cdot\nabla\times
(\mathbf{v}_{e}-\mathbf{v}_{i}).
\end{gather}
Using Eqs. (\ref{ni}) and (\ref{ne}), and eliminating
$\nabla\times (\mathbf{v}_{e}-\mathbf{v}_{i})$ with the aid of the
Maxwell equation (we neglect the displacement current for the
low-frequency motion)
\begin{equation}
\label{max1}
 \nabla\times\delta\mathbf{B}=\frac{4\pi en_{0}}{c}(\mathbf{v}_{i}-\mathbf{v}_{e}),
\end{equation}
one can obtain
\begin{gather}
 \frac{\partial^{2}\delta n}{\partial
t^{2}}-v_{s}^{2}\Delta \delta n-\frac{n_{0}}{B_{0}}v_{A}^{2}\Delta
\delta B=\frac{mn_{0}}{M}\nabla\cdot\mathbf{F}_{1}\nonumber \\
-\frac{m}{M}\nabla\cdot \frac{\partial\mathbf{F}_{2}}{\partial t},
\label{low1}
\end{gather}
where $v_{A}=B_{0}/\sqrt{4\pi n_{0}M}$ is the Alfv\'{e}n speed.
The second term in the right-hand side of Eq. (\ref{low1}) is
small compared to the first one by the factor
$\sim\omega/\omega_{UH}$ and can be neglected. To obtain equation
for the low-frequency magnetic field perturbation $\delta B$ we
substract Eq. (\ref{ve}) from Eq. (\ref{vi}) and take the $\curl$
\begin{gather}
\frac{\partial}{\partial t}\nabla\times
(\mathbf{v}_{i}-\mathbf{v}_{e})=\frac{e}{m}\nabla\times\mathbf{E}-
\omega_{ci}\hat{\mathbf{z}}\,\nabla\cdot\mathbf{v}_{i}\nonumber \\
 -
\omega_{ce}\hat{\mathbf{z}}\,\nabla\cdot\mathbf{v}_{e}+\nabla\times\mathbf{F}_{1}.
\end{gather}
Using Eqs. (\ref{ni}),(\ref{ne}), (\ref{max1}) and the Maxwell
equation
\begin{equation}
\nabla\times\mathbf{E}=-\frac{1}{c}\frac{\partial\delta\mathbf{B}}{\partial
t }
\end{equation}
we get
\begin{equation}
\label{low2} \frac{\partial}{\partial
t}\left(1-\frac{c^{2}}{\omega_{pe}^{2}}\Delta\right)\delta B-
\frac{B_{0}}{n_{0}}\frac{\partial\delta n}{\partial t}=\frac{4\pi
c}{\omega_{pe}^{2}}(\nabla\times\mathbf{F}_{1})_{z}.
\end{equation}
Representing
\begin{equation}
\mathbf{v}_{e}^{H}=\frac{1}{2}[\mathbf{v}\exp(-i\omega_{UH}t)+\mathrm{c.c.}],
\label{v1}
\end{equation}
from the high-frequency momentum equation for electrons we have
\begin{equation}
\mathbf{v}=\frac{e}{m}\frac{[i\omega_{UH}\nabla\varphi+
\omega_{ce}(\nabla\varphi\times\hat{\mathbf{z}})]}{(\omega_{UH}^{2}-\omega_{ce}^{2})}.
\label{v2}
\end{equation}
With the aid of the Maxwell equation for $\partial
\mathbf{B}^{H}/\partial t$, two terms in the expression for
$\mathbf{F}_{1}$ can be combined to yield
\begin{gather}
\mathbf{F}_{1}=\langle
(\mathbf{v}_{e}^{H}\cdot\nabla)\cdot\mathbf{v}_{e}^{H}+
[\mathbf{v}_{e}^{H}\times[\nabla\times\mathbf{v}_{e}^{H}]]\rangle
\nonumber \\
=\frac{1}{2}\langle\nabla\,(\mathbf{v}_{e}^{H}\cdot\mathbf{v}_{e}^{H})\rangle.
\label{F1}
\end{gather}
Using Eqs. (\ref{low1}), (\ref{low2})--(\ref{F1}) we obtain
equations for the low-frequency plasma density and magnetic field
perturbations
\begin{gather}
 \left(\frac{\partial^{2}}{\partial
t^{2}}-v_{s}^{2}\Delta\right)\frac{\delta n}{n_{0}}
-v_{A}^{2}\Delta \frac{\delta B}{B_{0}}=\frac{1}{16\pi n_{0} M }
 \nonumber \\
 \times \Delta\left\{\left(1+2\frac{\omega_{ce}^{2}}{\omega_{pe}^{2}}\right)
 |\nabla\varphi|^{2} \right. \nonumber \\
 \left. +2i\left(1+\frac{\omega_{ce}^{2}}{\omega_{pe}^{2}}\right)
 \frac{\omega_{ce}}
 {\omega_{UH}}[\nabla\varphi\times\nabla\varphi^{\ast}]_{z}\right\},
 \label{lo1}
\end{gather}

\begin{equation}
 \left(1-\frac{c^{2}}{\omega_{pe}^{2}}\Delta\right)\frac{\delta
B}{B_{0}}=\frac{\delta n}{n_{0}}.
 \label{lo2}
\end{equation}
In the linear approximation, Eqs. (\ref{lo1}) and (\ref{lo2}) give
the dispersion relation for the dispersive fast magnetosonic wave
\begin{equation}
\label{fast}
\Omega_{k}^{2}=k_{\perp}^{2}v_{s}^{2}+\frac{k_{\perp}^{2}v_{A}^{2}}
{1+k_{\perp}^{2}c^{2}/\omega_{pe}^{2}}.
\end{equation}
In what follows we omit the subscript $\perp$ in $k_{\perp}$.
Equations (\ref{eq1}), (\ref{lo1}) and (\ref{lo2}) form a closed
system of 2D equations describing the interaction between
upper-hybrid waves and dispersive magnetosonic waves. In the 1D
case these equations coincide with those obtained in Refs. $4$,
$5$ and $7$.

\section{nonlinear dispersion relation}
\label{sec3}
 In this section we consider the linear theory of the
modulational instability of a pump wave with a frequency close to
the upper-hybrid frequency. As usual, we express the low-frequency
perturbation of the plasma density as
\begin{gather}
\frac{\delta
n}{n_{0}}=\hat{n}\exp(i\mathbf{k}\cdot\mathbf{r}-i\Omega
t)+\mathrm{c}. \mathrm{c}. , \\
\frac{\delta
B}{B_{0}}=\hat{b}\exp(i\mathbf{k}\cdot\mathbf{r}-i\Omega
t)+\mathrm{c}. \mathrm{c}. ,
\end{gather}
while the upper-hybrid wave is decomposed into the pump wave and
two sidebands
\begin{gather}
\varphi=\varphi_{0}e^{i(\mathbf{k}_{0}\cdot\mathbf{r}-\delta_{0}
t)}+\varphi_{+}e^{i[(\mathbf{k}_{0}+\mathbf{k})\cdot\mathbf{r}-(\delta_{0}+\Omega)t]}
\nonumber \\
+\varphi_{-}e^{i[(\mathbf{k}_{0}-\mathbf{k})\cdot\mathbf{r}-(\delta_{0}-\Omega)t]}+
\mathrm{c}. \mathrm{c}.,
\end{gather}
where $\delta_{0}=\omega_{UH}k_{0}^{2}R^{2}/2$. The amplitudes of
the satellites can be calculated from Eq. (\ref{eq1}). We have
\begin{gather}
D_{+}\varphi_{+}=\alpha_{+}\hat{n}\varphi_{0}, \label{D1} \\
D_{-}\varphi_{-}^{\ast}=\alpha_{-}\hat{n}\varphi_{0}^{\ast},
\label{D2}
\end{gather}
where
\begin{gather}
\alpha_{\pm}=-(k_{0}^{2}\pm\mathbf{k}\cdot\mathbf{k}_{0})\left(\omega_{pe}^{2}+
\frac{2\omega_{ce}^{2}}{1+k^{2}\lambda_{e}^{2}}\right)\nonumber \\
+i(\mathbf{k}\times\mathbf{k}_{0})_{z}\frac{\omega_{ce}}{\omega_{UH}}
\left(\omega_{pe}^{2}+
\frac{\omega_{pe}^{2}+2\omega_{ce}^{2}}{1+k^{2}\lambda_{e}^{2}}\right),
\end{gather}
and the function
\begin{equation}
D_{\pm}=2\omega_{UH}(\mathbf{k}_{0}\pm\mathbf{k})^{2}(\delta_{\pm}\mp\Omega)
\end{equation}
is the Fourier transform of the linear operator in the left-hand
side of Eq. (\ref{eq1}) evaluated at
$\mathbf{k}_{0}\pm\mathbf{k}$, and
$\delta_{\pm}=\omega_{UH}R^{2}[(\mathbf{k}_{0}\pm\mathbf{k})^{2}-k_{0}^{2}]/2$
are the mismatches between the satellite frequencies and the
frequency of the pump wave.
 The amplitudes of the low-frequency perturbations are found
from Eqs. (\ref{lo1}) and (\ref{lo2}):
\begin{equation}
\label{D3}
 (\Omega^{2}-\Omega_{k}^{2})\hat{n}=\frac{k^{2}(\beta_{+}\varphi_{+}\varphi_{0}^{\ast}+
\beta_{-}\varphi_{-}^{\ast}\varphi_{0})}{16\pi n_{0}M},
\end{equation}
\begin{equation}
(1+k^{2}c^{2}/\omega_{pe}^{2})\hat{b}=\hat{n},
\end{equation}
where
\begin{gather}
\beta_{\pm}=\left(1+2\frac{\omega_{ce}^{2}}{\omega_{pe}^{2}}\right)
(k_{0}^{2}\pm\mathbf{k}\cdot\mathbf{k}_{0}) \nonumber \\
+2i\left(1+\frac{\omega_{ce}^{2}}{\omega_{pe}^{2}}\right)(\mathbf{k}\times\mathbf{k}_{0})_{z}
\end{gather}
and $\Omega_{k}^{2}$ is determined by Eq. (\ref{fast}).
 By combining Eqs. (\ref{D1}), (\ref{D2}) and (\ref{D3}) one can
obtain a nonlinear dispersion relation
\begin{equation}
\label{disp}
 \Omega^{2}-
\Omega^{2}_{k}=\frac{k^{2}|\varphi_{0}|^{2}}{16\pi
n_{0}M}\left(\frac{\alpha_{+}\beta_{+}}{D_{+}}+\frac{\alpha_{-}\beta_{-}}{D_{-}}\right).
\end{equation}
 Equation (\ref{disp}) generalizes the nonlinear dispersion relation
obtained by Eliasson and Shukla \cite{Shukla3} by considering 2D
case and including the vector nonlinearity. For 1D case we recover
the previous result. Note, that in the case of coplanar (in the
plane perpendicular to the magnetic field) wave vectors
$\mathbf{k}\parallel\mathbf{k}_{0}$, the parametric coupling of
the waves due to the vector nonlinearity is absent, while the
coupling due to the scalar nonlinearity is the most effective. In
the opposite case, i.e. $\mathbf{k}\perp\mathbf{k}_{0}$, the
interaction due to the vector nonlinearity is the most effective,
while the interaction due to the scalar nonlinearity is the least
effective (and absent for $k_{0}\ll k$). In general case
$\mathbf{k}\nparallel\mathbf{k}_{0}$, and when
$\omega_{pe}\sim\omega_{ce}$ so that both types of the
nonlinearities yield comparable contribution, this leads to a
rather complicated picture of the parametric instability. In this
paper, we restrict ourselves to the case of a weakly magnetized
plasma with $\omega_{ce}^{2}/\omega_{pe}^{2}\ll 1$. Results,
concerning the case of a moderately magnetized plasma with
$\omega_{ce}\sim\omega_{pe}$, including the nonlinear analysis
(see below),
 will be published elsewhere.

In the case $\omega_{ce}\ll\omega_{pe}$, one can neglect the
vector nonlinearity and Eq. (\ref{disp}) takes the form
\begin{equation}
\label{dis2}
\Omega^{2}_{k}-\Omega^{2}=\frac{|E_{0}|^{2}k^{2}\omega_{pe}}{32\pi
n_{0} M}\left[\frac{\cos^{2}\mu_{+}}
{(\delta_{+}-\Omega)}+\frac{\cos^{2}\mu_{-}} {(\delta_{-}+\Omega)}
\right],
\end{equation}
where
\begin{equation}
|E_{0}|^{2}=k_{0}^{2}|\varphi_{0}|^{2} , \quad \cos
\mu_{\pm}=\frac{\mathbf{k}_{0}\cdot
(\mathbf{k}_{0}\pm\mathbf{k})}{k_{0}|\mathbf{k}_{0}\pm\mathbf{k}|}.
\end{equation}
When both sidebands are resonant and the low-frequency
perturbations are nonresonant, we can consider the limiting case
$\mathbf{k}_{0}\gg\mathbf{k}$ and $\Omega_{k}\gg\Omega$. Then, the
dispersion relation Eq. (\ref{dis2}) takes the form
\begin{equation}
\label{dis3}
(\Omega-\mathbf{k}\cdot\mathbf{v}_{g})^{2}=\frac{\omega_{pe}^{2}k^{4}R^{4}}{4}
\left[1-\frac{|E_{0}|^{2}}{8\pi n_{0}MR^{2}\Omega_{k}^{2}}\right],
\end{equation}
where $\mathbf{v}_{g}$ is the group velocity of the UH wave.
Equation (\ref{dis3}) predicts the instability when
$|E_{0}|^{2}>8\pi n_{0}MR^{2}\Omega_{k}^{2}$. In the opposite case
of long-wavelength pump $\mathbf{k}\gg\mathbf{k}_{0}$ Eq.
(\ref{dis2}) is reduced to
\begin{equation}
\label{dis4}
\Omega^{2}=\frac{1}{2}\left\{\delta^{2}+\Omega_{k}^{2}\pm
\sqrt{(\delta^{2}-\Omega_{k}^{2})^{2}+\frac{|E_{0}|^{2}\omega_{pe}k^{2}\delta}{4\pi
n_{0}M}}\right\},
\end{equation}
where $\delta=\omega_{UH}k^{2}R^{2}/2$ and, thus, there is a
purely growing instability for sufficiently large amplitudes of
the pump wave $E_{0}$.

\section{Proof of absence of collapse}
\label{sec4}

In what follows, we will consider the case
$\omega_{ce}^{2}/\omega_{pe}^{2}\ll 1$, so that vector
nonlinearities in Eqs. (\ref{eq1}) and (\ref{lo1}) can be
neglected. Then, introducing the dimensionless variables
\begin{gather}
t\rightarrow \omega_{LH}t, \quad \mathbf{r}\rightarrow
\mathbf{r}\,\omega_{LH}/v_{s},\\
\varphi\rightarrow \varphi\frac{\omega_{pi}}{4v_{s}\sqrt{\pi
n_{0}T_{e}}}, \quad b\rightarrow \frac{\delta
B}{B_{0}}\frac{\omega_{pe}^{2}}{\omega_{ce}^{2}\beta},\\
\mu=\frac{2\omega_{UH}m}{3\omega_{LH}M}, \quad
 \beta=\frac{v_{s}^{2}}{v_{A}^{2}},
\end{gather}
and eliminating $\delta n/n_{0}$, one can rewrite Eqs.
(\ref{eq1}), (\ref{lo1}) and (\ref{lo2}) as follows
\begin{equation}
\label{main1} \Delta \left(i\mu\frac{\partial\varphi}{\partial
t}+\Delta\varphi\right)=\nabla\cdot[(\beta b-\Delta
b)\nabla\varphi],
\end{equation}
\begin{equation}
\label{main2} \left(\frac{\partial^{2}}{\partial
t^{2}}-\Delta\right)(\beta-\Delta)b -\Delta
b=\Delta|\nabla\varphi|^{2},
\end{equation}
Equations (\ref{main1}) and (\ref{main2}) conserves energy
\begin{equation}
\label{en} N=\int |\nabla\varphi|^{2}\,d\mathbf{r},
\end{equation}
and Hamiltonian
\begin{equation}
\label{ham} H=\int
\left(|\Delta\varphi|^{2}+\frac{1}{2}|\nabla\varphi|^{2}(\beta-\Delta)b\right)\,d\mathbf{r},
\end{equation}
and can be written in the Hamiltonian form
\begin{equation}
-i\mu\Delta\frac{\partial\varphi}{\partial t}=\frac{\delta
H}{\delta\varphi^{\ast}}.
\end{equation}

In this section, following the ideas suggested in Ref. $14$, we
present a rigorous proof of the absence of collapse for the
stationary 2D solutions of the form
$\varphi(x,y,t)=\psi(x,y)\exp(i\Lambda t)$ in the model described
by Eqs. (\ref{main1}) and (\ref{main2}). We use an exact approach
\cite{ZaharovKuznezov} based on the Liapunov stability theory. Let
us briefly recall the essence of the Liapunov method. For an
invariant set of the dynamical system (in particular, for the set
of stationary solutions $\{u_{s}\}$) to be stable, it is
sufficient that there exists a functional $L[u]$ with the
following properties: a) $L$ is the positive definite functional
for the perturbed states $\{u\}$, i.e. $L[u]\geqslant 0$; b) $L$
reaches its minimum on the set $\{u_{s}\}$, $L[u_{s}]=0$; c) $L$
is nonincreasing function of time $t$, i.e. $dL/dt\leqslant 0$.
For hamiltonian systems, these conditions are equivalent to the
requirement that the Hamiltonian is bounded from below under the
fixed conserved quantity $N$ (then, one can choose $L=H-H_{min}$).
Then, the 2D stationary localized solutions corresponding to the
global minimum of $H$ (fundamental solitons) are stable in the
Liapunov sense.

Note, that for the stationary solutions, Eqs. (\ref{main1}) and
(\ref{main2}) can rewritten as follows
\begin{equation}
\label{seq1} \Delta
(-\Lambda\mu+\Delta)\psi=\nabla\cdot[(\beta-\Delta)b\nabla\psi],
\end{equation}
with
\begin{equation}
\label{seq2}
 b(\mathbf{r})=-\int
G(\mathbf{r}-\mathbf{r}')|\nabla\varphi(\mathbf{r}')|^{2}\,d\mathbf{r}',
\end{equation}
where $G(\mathbf{r})=K_{0}(\sqrt{\beta+1}|\mathbf{r}|)/2\pi$ is
the Green function satisfying the equation
$(\beta+1-\Delta)G(\mathbf{r})=\delta(\mathbf{r})$, and $K_{0}(z)$
is the modified Bessel function of the second kind of order zero.
It is seen, that the nonlinearity in Eq. (\ref{seq1}) has
essentially nonlocal character. We represent the Hamiltonian
(\ref{ham}) as
\begin{equation}
\label{ham1} H=A+\frac{1}{2}P,
\end{equation}
where $A=\int |\Delta\psi|^{2}\,d\mathbf{r}$ and the nonlinear
term $P$ can be written as
\begin{gather}
P=-\int
|\nabla\varphi(\mathbf{r})|^{2}|\nabla\varphi(\mathbf{r}')|^{2}(\beta-\Delta)
G(|\mathbf{r}-\mathbf{r}'|) \,d\mathbf{r}\,d\mathbf{r}' \nonumber \\
=-\int |\nabla\varphi(\mathbf{r})|^{4}\,d\mathbf{r}+C,
\end{gather}
where
\begin{equation}
C=\int
|\nabla\varphi(\mathbf{r})|^{2}|\nabla\varphi(\mathbf{r}')|^{2}
G(|\mathbf{r}-\mathbf{r}'|)\,d\mathbf{r}\,d\mathbf{r}'>0.
\end{equation}
To get an estimate for $C$ we use the following inequality
\cite{Tur}
\begin{equation}
\label{tur}
\int\frac{f^{2}(\mathbf{r})}{|\mathbf{r}-\mathbf{r'}|}\,d\mathbf{r}
\leqslant 2\left(\int
f^{2}\,d\mathbf{r}\right)^{1/2}\left(\int(\nabla
f)^{2}\,d\mathbf{r}\right)^{1/2},
\end{equation}
where $f(\mathbf{r})$ is an arbitrary  sufficiently smooth
function, the integration in Eq. (\ref{tur}) is performed over
entire 2D space. We have
\begin{gather}
\label{chain2} C=\int
|\nabla\varphi(\mathbf{r})|^{2}\frac{|\nabla\varphi(\mathbf{r}')|^{2}}
{|\mathbf{r}-\mathbf{r}'|}
G(|\mathbf{r}-\mathbf{r}'|)|\mathbf{r}-\mathbf{r}'|\,d\mathbf{r}\,d\mathbf{r}'\nonumber
\\
\leqslant M\int |\nabla\varphi(\mathbf{r})|^{2}d\mathbf{r}\int
\frac{|\nabla\varphi(\mathbf{r}')|^{2}}
{|\mathbf{r}-\mathbf{r}'|}d\mathbf{r}' \nonumber \\
\leqslant 2M \left(\int
|\nabla\varphi(\mathbf{r})|^{2}d\mathbf{r}\right)^{3/2} \left(\int
|\Delta\varphi(\mathbf{r})|^{2}d\mathbf{r}\right)^{1/2}\nonumber
\\
=2MN^{3/2}A^{1/2},
\end{gather}
where $M=\max_{z}[z K_{0}(z\sqrt{\beta+1})/2\pi]>0$ and we have
used Eq. (\ref{tur}). Next we use the inequality
\begin{equation}
\int |\nabla\varphi(\mathbf{r})|^{4}\,d\mathbf{r}\leqslant N^{2}.
\end{equation}
As a result, one can write the obvious chain of inequalities:
\begin{equation}
P\geqslant -N^{2}+C\geqslant -N^{2}-C\geqslant
-N^{2}-2MN^{3/2}A^{1/2}.
\end{equation}
 Inserting this estimate into the expression for the
hamiltonian Eq. (\ref{ham1}) we get
\begin{equation}
\label{H4}
 H\geqslant A-\frac{1}{2}N^{2}-MN^{3/2}A^{1/2}.
\end{equation}
Under the fixed $N$, the right hand side of the inequality
(\ref{H4}) reaches its minimum at $A=M^{2}N^{3}/4$, so that
\begin{equation}
\label{H5}
 H\geqslant -M^{2}N^{3}/4-N^{2}/2.
\end{equation}
Thus, we have showed that, under the fixed conserved quantity $N$,
the Hamiltonian is bounded from below. According to standard
Liapunov theory, this represents a rigorous proof that a collapse
with the wave amplitude locally going to infinity cannot occur for
the stationary solutions in the model described by Eqs.
(\ref{main1}) and (\ref{main2}).

\section{nonlinear localized solutions}
\label{sec5}

We consider the case when the upper-hybrid wavelength is much
smaller than the typical spatial scale of the localization and
represent (in the corresponding dimensionless variables)
\begin{equation}
\varphi=\Phi\exp(i\mathbf{k}_{0}\cdot\mathbf{r}-i\delta_{0}t).
\end{equation}
We assume that $k_{0}L\gg 1$, where $L=|\psi/\nabla\psi|$ is the
typical length scale of the low-frequency perturbation. We also
make an additional assumption of $\mathbf{k}_{0}\cdot\nabla=0$.
Under this, the group velocity of the upper-hybrid wave is
perpendicular to the direction of inhomogeneity and there is no
transportation of energy in the direction of inhomogeneity which
results in stationary envelope structure in the direction
perpendicular to the magnetic field.
 Then, the
system Eqs. (\ref{main1}) and (\ref{main2}) takes the form
\begin{equation}
\label{m1} i\mu\frac{\partial\Phi}{\partial
t}+\Delta\Phi=(\beta-\Delta)b\Phi,
\end{equation}
\begin{equation}
\label{m2} \left(\frac{\partial^{2}}{\partial
t^{2}}-\Delta\right)(\beta-\Delta)b -\Delta b=\Delta|\Phi|^{2},
\end{equation}
where we have rescaled $\Phi$ so that $\Phi\rightarrow k_{0}\Phi$.
The proof of the absence of collapse for the stationary solutions
of the system (\ref{m1}) and (\ref{m2}) is quite analogues to that
presented in the previous section.

We look for stationary solutions of Eqs. (\ref{m1}) and (\ref{m2})
in the form $\Phi(x,y,t)=\Psi(x,y)\exp(i\lambda t/\mu)$, where
$\lambda/\mu$ is the nonlinear frequency shift, so that $\Psi$
obeys the equation
\begin{equation}
\label{m3} -\lambda\Psi+\Delta\Psi=(\beta-\Delta)b\Psi,
\end{equation}
\begin{equation}
\label{m4} (\beta+1-\Delta)b=-|\Psi|^{2}.
\end{equation}
To solve numerically Eqs. (\ref{m3}) and (\ref{m4}), we impose
periodic boundary conditions on Cartesian grid and use the
relaxation technique similar to one described in Ref. $17$.
Choosing an appropriate initial guess, one can find numerically
with high accuracy (the norms of the residuals were less than
$10^{-9}$) three different classes of spatially localized
solutions of Eqs. (\ref{m3}) and (\ref{m4}) -- the nonrotating
(multi)solitons, the radially symmetric vortices, and the rotating
multisolitons (azimuthons).

\begin{figure}
\begin{center}\includegraphics[width=3.4in]{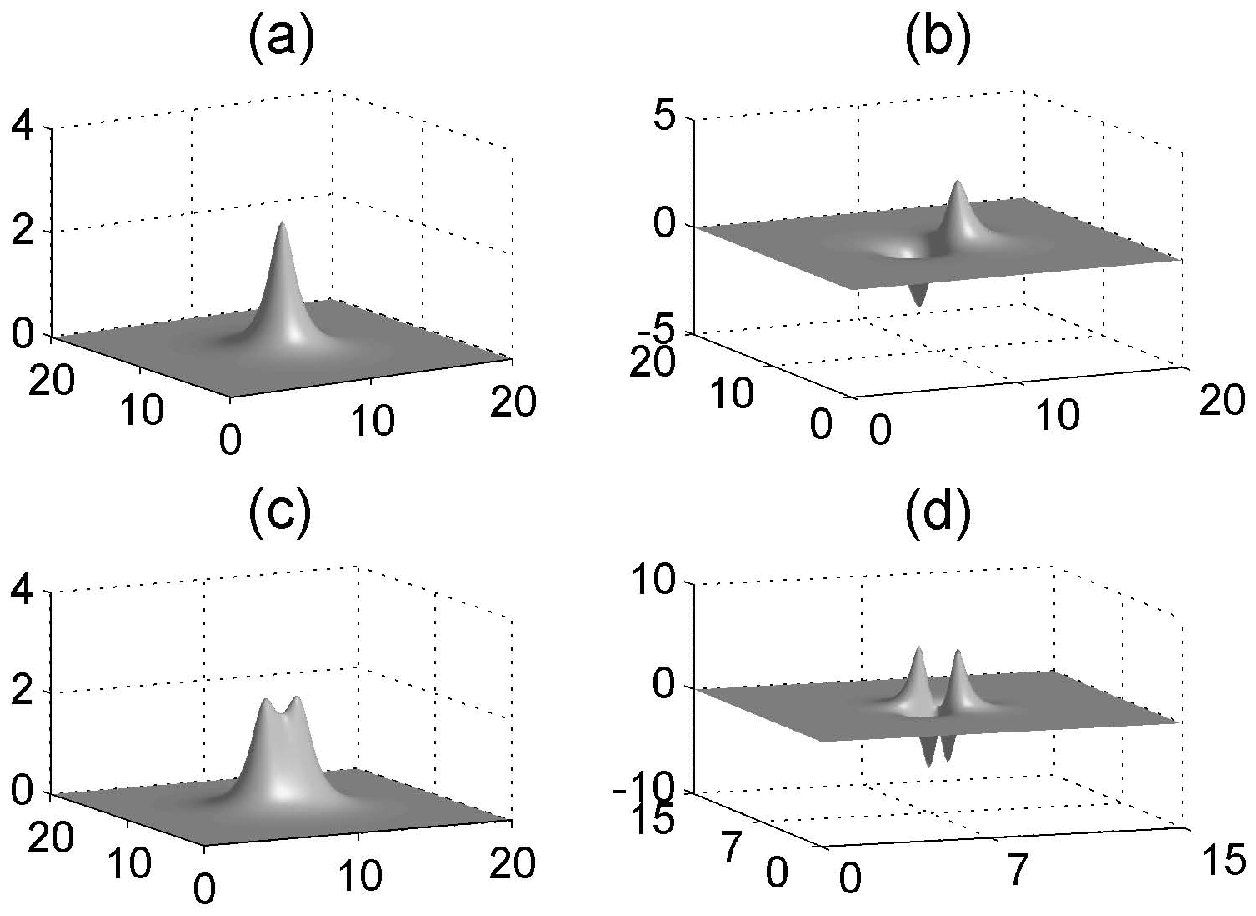}\end{center}
\caption{\label{fig1} Numerically found nonrotating stationary
localized solutions of Eqs. (\ref{m1}) and (\ref{m2}) for
$\beta=0.1$: (a) monopole with $\lambda=0.5$; (b) dipole with
$\lambda=0.5$; (c) two-hump soliton with $\lambda=0.5$; (d)
quadrupole with $\lambda=2$. The real part of the field $\Psi$ is
shown. }
\end{figure}

The real (or containing only a constant complex factor) function
$\Psi(x,y)$ corresponds to nonrotating solitary structures.
Examples of such nonrotating (multi)solitons for Eqs. (\ref{m3})
and (\ref{m4}), namely, a monopole, a dipole, two-hump soliton,
and a quadrupole are presented in
Figs.~\ref{fig1}(a)-~\ref{fig1}(d) for the case $\beta=0.1$. The
analogous solutions can be found for $\beta>1$. The nonrotating
multipoles consist of several fundamental solitons (monopoles)
with opposite phases.

\begin{figure}
\includegraphics[width=3.4in]{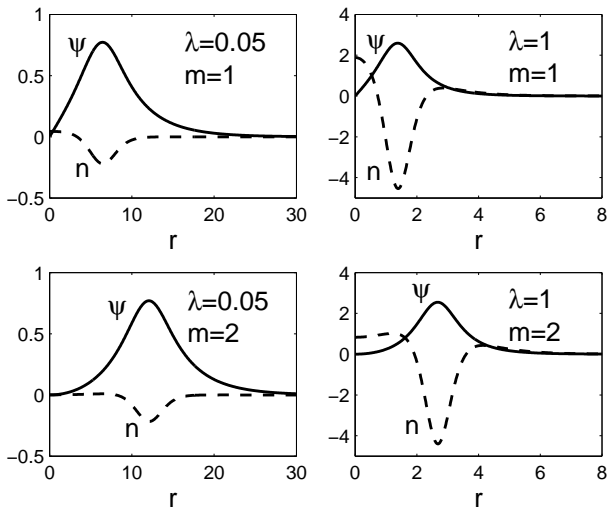}
\caption{\label{fig2} Examples of the stationary radially
symmetric vortex solutions, characterized by the topological
charges $m=1,2$ for $\lambda=0.05$ (left panel) and $\lambda=1$
(right panel): radial profiles of the field intensity $|\Psi|$
(solid curve) and density $n$ (dashed curve) are shown for
$\beta=0.25$. }
\end{figure}

The second class of solutions, vortex solutions, are the solutions
with the radially symmetric amplitude $|\Psi(x,y)|$, that vanishes
at the center, and a rotating spiral phase in the form of a linear
function of the polar angle $\theta$, i.e. $\arg\Psi=m\theta$,
where $m$ is an integer. The index $m$  (topological charge)
stands for a phase twist around the intensity ring. The important
integral of motion associated with this type of solitary wave is
the $z$-component of the angular momentum
\begin{equation}
M_{z}= \mathrm{Im}\,\int
\left[\Phi^{\ast}(\mathbf{r}\times\nabla_{\perp}\Phi)\right]_{z} d
\textbf{r},
\end{equation}
which can be expressed through the soliton amplitude $U$ and phase
$\phi$,
\begin{equation}
M_{z}= \int \frac{\partial\phi}{\partial\theta}U^{2} d \textbf{r},
\end{equation}
and for the vortex we have $M_{z}=mN$. Examples of the stationary
radially symmetric vortex solutions, characterized by the
topological charges $m=1,2$ for different values of $\lambda$ are
shown in Fig.~\ref{fig2}.

\begin{figure}
\includegraphics[width=3.4in]{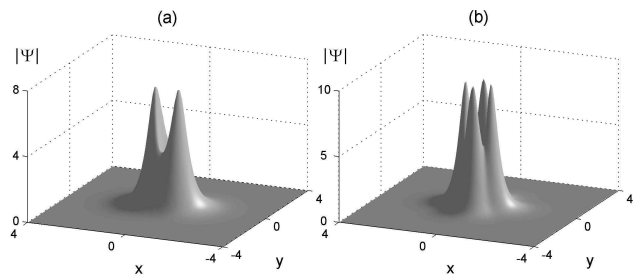}
\caption{\label{fig3} Examples of the rotating multisolitons
(azimuthons) with $\lambda=5$ for $\beta=0.1$: a) azimuthon with
two intensity peaks; b) azimuthon with four intensity peaks. The
field intensity $|\Psi|$ is shown.}
\end{figure}

The third class of solutions, rotating multisolitons with the
spatially modulated phase, were first introduced in Ref. $18$ for
models with local nonlinearity, where they were called azimuthons.
The azimuthons can be viewed as an intermediate kind of solutions
between the rotating radially symmetric vortices and nonrotating
multisolitons. Using variational analysis to describe azimuthons,
the authors of Ref. $12$ considered the following trial function
in polar coordinates ($r$,$\theta$)
\begin{equation}
\label{trial} \Psi(r,\theta)=r^{|m|}\Phi(r)(\cos m\theta+ip\,\sin
m\theta),
\end{equation}
where $\Phi$ is the real function, which vanish fast enough at
infinity, $m$ is an integer, and $0\leq p\leq 1$. The case $p=0$
corresponds to the nonrotating multisolitons (e. g. $m=1$ to a
dipole, $m=2$ to a quadrupole etc.), while the opposite case $p=1$
corresponds to the radially symmetric vortices. The intermediate
case $0<p<1$ corresponds to the azimuthons. The trial function Eq.
(\ref{trial}) was chosen as an initial guess in our numerical
relaxation method. Then, the numerically found complex function
$\Psi(x,y)$ with a spatially modulated phase corresponds to the
azimuthons.  We introduced the parameter $p$ (modulational depth),
which is similar to the one in Eq. (\ref{trial}), in the following
way
\begin{equation}
p=\max|\mathrm{Im}\,\Psi|/\max|\mathrm{Re}\,\Psi|.
\end{equation}
For fixed $\lambda$, there is a family of azimuthons with
different $p$. Since the azimuthons have a nontrivial phase, they,
like the radially symmetric vortices, carry out the nonzero
angular momentum. In Figures~\ref{fig3} we demonstrate two
numerically found examples of the azimuthons with two and four
intensity peaks for the nonlocal model described by Eqs.
(\ref{m3}) and (\ref{m4}). The azimuthon with two intensity peaks
consists of two dipole-shaped structures in the real and imaginary
parts of $\Psi$ with different amplitudes and the ratio of these
amplitudes is the modulational depth $p$. Note, that the choice of
initial guess in the relaxation method for finding the azimuthons
is much more sophisticated than that for the nonrotating
multisolitons or vortices. For example, we were not able to find
azimuthon solutions with predetermined (in advance) value of $p$.

We next addressed the stability of these localized solutions and
study the evolution of the solitons in the presence of small
initial perturbations. We have undertaken extensive numerical
modeling of Eqs. (\ref{m1}) and (\ref{m2}) initialized with our
computed solutions with added gaussian noise. The initial
condition was taken in the form $\Psi(x,y)[1+\varepsilon f(x,y)]$,
where $\Psi(x,y)$ is the numerically calculated exact solution,
$f(x,y)$ is the white gaussian noise with variance $\sigma^{2}=1$
and the parameter of perturbation $\varepsilon=0.005 \div 0.01$.
In addition, azimuthal perturbation of the form
$i\varepsilon\sin\theta$ was taken for the vortices and
azimuthons. Spatial discretization was based on the pseudospectral
method. Temporal $t$-discretization included the split-step
scheme.

\begin{figure}
\includegraphics[width=3.4in]{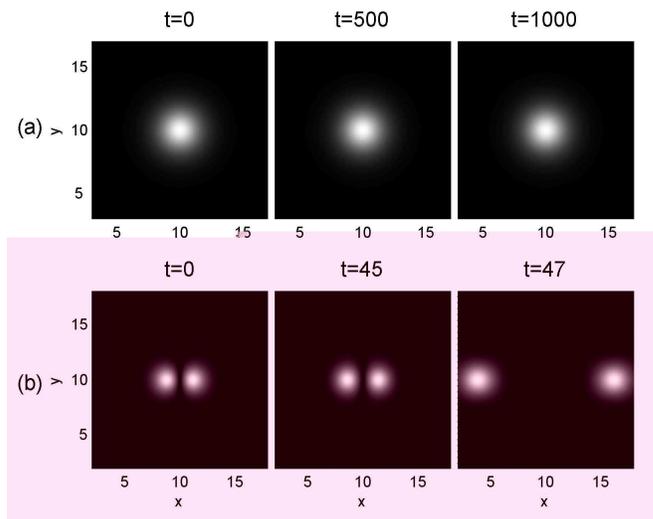}
\caption{\label{fig4}  (a) Stable evolution of the monopole with
$\lambda=1$; (b) splitting of the dipole with $\lambda=2$ into two
monopoles. }
\end{figure}

Numerical simulations clearly show that the fundamental solitons
are stable and do not collapse even for the negative initial
hamiltonian. Stable evolution of the monopole soliton with
$\lambda=1$ is shown in Fig.~\ref{fig4}(a). We have observed
neither stable evolution nor collapse for multisolitons. If the
nonlinear frequency shift $\lambda$ is not too large, the
multisolitons decay into several monopole solitons, but can
survive over quite considerable times. Splitting of the dipole
soliton in two monopoles which move in the opposite directions
without changing their shape is shown in Fig.~\ref{fig4}(b).
Figure 5 presents an example of the decay of the vortex into three
fundamental solitons. Since the total angular momentum is
conserved, the monopole solitons fly off the ring along tangential
trajectories. A similar behavior was observed for the rotating
multisolitons (azimuthons with two and four intensity peaks). If
the nonlinear frequency shift $\lambda$ exceeds some critical
value depending on the multisoliton type (i.e. dipole, azimuthon
with two intensity peaks etc.), the unstable multisoliton turn
into the one monopole with larger amplitude. Although the
multisolitons turn out to be unstable in all our runs, one cannot
exclude the possibility of existence of stable multisolitons in
some narrow range of parameters (see, for example, Refs. $19$ and
$20$). A rigorous proof of the stability/instability could give a
linear stability analysis with the corresponding eigenvalue
problem.

Soliton structures can arise as the result of modulational
instability of the initial monochromatic UH pump wave. We
numerically studied the evolution of a monochromatic long
wavelength UH wave in the framework of Eqs. (\ref{m1}) and
(\ref{m2}). Consider the case
 when a monochromatic UH wave of the form
\begin{equation}
\label{homo} \Phi(x,y)=1.5\,\exp(0.07ix+0.07iy)
\end{equation}
is chosen as an initial condition in a box of side $L=10$. Thus,
the wave Eq. (\ref{homo}) represents the almost homogeneous
initial field. The time evolution of the field Eq. (\ref{homo}) is
shown in Fig. \ref{fig6}. It is clearly seen the formation of one
well-shaped soliton, which moves in some direction with almost no
change in shape and at later times coexists with turbulent
environment. Depending on the amplitude and wave vector of the
pump as well as on the box length, we observed the formation of
two and more solitons, and even, in some cases, the formation
(simultaneously with solitons) of structures resembling the
ring-like vortices. No soliton formation, however, was detected in
all cases if the amplitude of the initial pump was sufficiently
small (for example, less than $0.8$ for Eq. (\ref{homo})) so that,
for a given grid, box size and the wave vector of the pump, the
threshold of the modulational instability was not exceeded. It is
important, that in all our numerical simulations of the evolution
of initial fields we did not observed any evidence of collapse.

\begin{figure}
\includegraphics[width=3.4in]{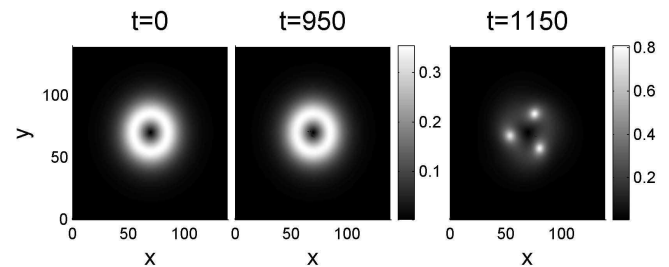}
\caption{\label{fig5} Evolution of the vortex with $m = 1$ and
$\lambda = 0.01$ for $\beta = 0.25$.}
\end{figure}

\begin{figure}
\includegraphics[width=3.4in]{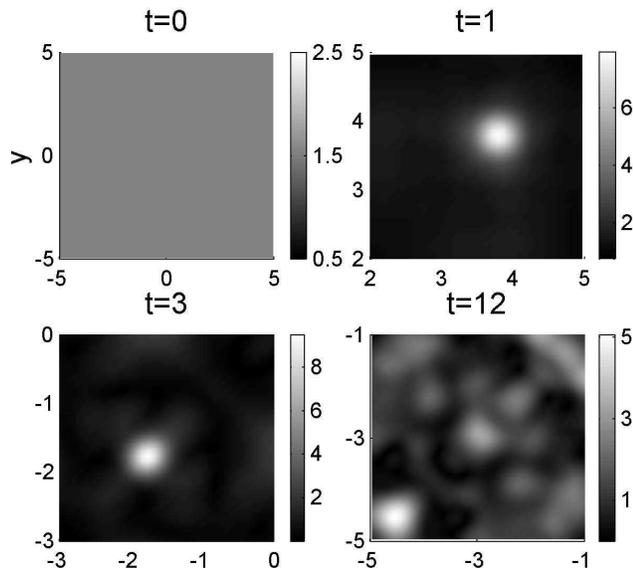}
\caption{\label{fig6} Evolution of the almost homogeneous initial
field Eq. (\ref{homo}) in the model Eqs. (\ref{m1}) and
(\ref{m2}). The panels show the intensity $|\Phi|$ in the $(x,y)$
plane at times $t=0$, $t=1$, $t=3$, and $t=12$, respectively. The
modulational instability of the initial field results in emergence
of a self-localized state (soliton).}
\end{figure}

\section{Conclusion}
\label{sec6}

In the present paper, we have considered the nonlinear interaction
between dispersive magnetosonic and high-frequency UH waves ion 2D
geometry. We have derived a set of 2D equations describing the
dynamics of nonlinearly coupled magnetosonic and UH waves.
Nonlocal nonlinearity in the equations prevents collapse and
results in the possibility of existence of stable 2D nonlinear
structures. We have presented a rigorous proof of the absence of
collapse in the model under consideration. We have found
numerically different types of nonlinear 2D localized structures
such as fundamental solitons, radially symmetric vortices,
nonrotating multisolitons (dipoles and quadrupoles), and rotating
multisolitons (azimuthons). By direct numerical simulations we
have  shown that fundamental solitons are stable and do not
collapse even with negative initial hamiltonian. Multisolitons and
vortices decay into fundamental solitons but can survive over
quite considerable times.

\end{document}